\documentstyle[preprint,revtex,eqsecnum]{aps}
\def\ubar{\bar u}

\def\g{\gamma}
\def\a{\alpha}
\def\m{\mu}
\def\b{\beta}
\def\psla{\not\!p}
\def\qsla{\not\!q}
\def\ksla{\not\!k}
\def\eq{$e^-q \rightarrow e^+ 2\mu^- q$}
\def\ee{$e^+e^-\rightarrow 2e^+ 2\mu^-$ or $2e^- 2\mu^+$}
\def\ep{$e^- p \rightarrow e^+ \mu^- \mu^- + anything$}
\begin{document}
\draft
\preprint{IFP-428-UNC}
\preprint{June 1992}
\begin{title}
\begin{center}
Dilepton Production in $e^- p$ and
$e^+ e^-$ Colliders
\end{center}
\end{title}
\author{Jyoti Agrawal, Paul H. Frampton and Daniel Ng}
\begin{instit}
\begin{center}
Institute of Field physics\\
Department of Physics and Astronomy\\University of North Carolina\\
Chapel Hill, North Carolina 27599-3255
\end{center}
\end{instit}
\begin{abstract}
%{\centerline \bf ABSTRACT}
In an $e^- p$ collider, a striking signature
for a dilepton gauge boson is \ep \ ; this cross-section is
calculated by using the helicity amplitude technique.
At HERA, with center-of-mass energy $\sqrt s = 314 GeV$,
a dilepton mass above $150 GeV$ is inaccessible but at LEPII-LHC, with
a center-of-mass energy $\sqrt s = 1790 GeV $,
masses up to 650 GeV can be discovered. In an $e^+ e^-$ collider, the signature
is \ee \ .  The cross-sections of this process are also calculated for the
center-of-mass energies $\sqrt s = 200, 500$ and $1000 GeV$.
\end{abstract}

\section{INTRODUCTION}

In a recent paper\cite{1}, two of the present authors (P.H.F. and D.N.)
have studied the phenomenology of dilepton gauge bosons predicted by
certain simple extensions of the standard model of strong and electroweak
interactions. In particular, the existence of a $SU(2)_L$ doublet
$(X^{--},X^-)$
of vector gauge bosons with lepton number $L=2$ is a plausible prediction of
a general class of theories in which the electroweak $ SU(2) \otimes U(1)$
gauge group is expanded to $ SU(3) \otimes U(1)$
(e.g. Ref \cite{2}). The crucial
theoretical and practical question is then what is the mass scale $ M_X$ ?

In Ref. \cite{1}, a lower bound of $M_X > 120 GeV$ was established by studying
$e^+ e^- \rightarrow e^+ e^- $ as well as the "wrong" muon decay
$\mu^- \rightarrow
e^- \nu_e \bar \nu_\mu $ ; the s-channel resonance in $e^- e^-\rightarrow
X^{--}\rightarrow\mu^-\mu^-$ was also computed.
For polarized muons, a stronger
limit \cite{3} of $M_X > 230$ GeV was estimated. Since certain assumption
about the couplings of the dilepton were made in Refs.\cite{1,2} we shall here
entertain a more general range $100 GeV < M_X < 1 TeV$ , although it should be
borne in mind that the lower end is probably already excluded by existing
data.

In the present paper,
we shall focus on some striking signatures of lepton number
violating processes in electron-proton and electron-positron colliders.
A light dilepton gauge boson as anticipated in Ref. \cite{2} couples
democratically to the three lepton family associated with $e$,$\mu$ and $\tau$.
Total lepton number $L=L_e+L_\mu+L_\tau$ is conserved but the separate
flavors of lepton $L_e$, $L_\mu$, $L_\tau$ are violated.
This is different from the minimal standard model where $L_e$, $L_\mu$,
$L_\tau$ are necessarily separately conserved.  This in turn means that there
exist dramatic signatures for light (below $1TeV$) dileptons which violate
$L_e$, $L_\mu$ and hence have no background events from
standard model processes;
such evidence for a dilepton gauge boson will be accessible to the next
generation of $e^-p$ and $e^+e^-$ colliders as we shall show by
explicit estimates of the relevant cross sections.

In an electron-proton collider one may see the process
$e^-p\rightarrow e^+\mu^-\mu^- + anything$
with zero background from the Standard Model. This is relevant to the HERA
collider at DESY in Hamburg, Germany; presently beginning operation with $30
GeV$ electrons colliding on $820GeV$ protons $(\sqrt s =314GeV)$
with luminosity \cite{4} $1.6 \times 10^{31}cm^{-2}s^{-1}
(0.5 fb^{-1}yr^{-1})$. In the future an
$e^-p$ collider is planned at CERN with 100 GeV electrons on 8000 GeV protons
(LEPII-LHC) ($ \sqrt s = 1790 GeV$) and luminosity
$2\times 10^{32} cm^{-2} s^{-1} (6fb^{-1}yr^{-1})$.

In an $e^+ e^-$ collider, one may see the background-free process \ee \ .
This is relevant to LEPII and the Next Linear Collider (NLC) with
the center-of-mass energies $\sqrt s = 200, 500$ and $1000 GeV$ and
luminosities \cite{4} $ 1.7 \times 10^{31}, 1 \times 10^{32}$ and
$1 \times 10^{33} cm^{-2}s^{-1}$ ($0.5, 3$ and $30 fb^{-1}yr^{-1}$).

The outline of this paper is as follows.  In Sect. II, we compute the
amplitude of the Feynman diagrams of the above processes.  In Sect. III,
the cross-sections are calculated for $e^-p$ and $e^+e^-$
colliders.  In Sect. IV, there are some concluding remarks.
Appendix A contains the analysis of production of a real on-shell dilepton;
we used this to check our computations.

\section{Feynman diagrams and helicity amplitudes}

\subsection{Preliminary}
For the processes we consider in this paper,
it is most convenient to calculate Feynman diagrams using the method
of helicity amplitudes, particularly when the external particles are
taken to be massless, which is a sensible approximation in the present case.
The formalism can be found in Ref. \cite{5}.
The outer product of a massless spinor with momentum $p$ and helicity
$\lambda (= \pm 1)$ is
\begin{equation}
u_{\lambda}(p) \ubar _{\lambda}(p) = \omega_{\lambda} \psla
\;,\  \omega _\lambda = {1 \over 2}(1 + \lambda \gamma _5)\;.
\end{equation}
Let us define two four-vectors $k_0^\mu$ and $k_1^\mu$ with the following
properties:
\begin{equation}
k_0 \cdot k_0 = 0\; , \; k_1 \cdot k_1 = -1 \; , \; k_0 \cdot k_1 = 0 \; .
\end{equation}
Hence any massless spinors with momentum $p$ and helicity $\lambda$
can be constructed from $u_- (k_0)$ by the following relations,
\begin{equation}
u_+(k_0)=\not\!k_1 u_-(k_0) \; ,
\; u_\lambda (p)=\psla u_{-\lambda}(k_0) / \sqrt{2 p \cdot k_0}\;.
\end{equation}
The expressions in Eq.(2.3) can be verified by substituting into Eq.(2.1).
{}From the second equation, we have $u_\lambda (-p) = i u_\lambda (p)$.
Therefore, there is an (unobservable) overall phase when we replace an
antifermion spinor by a fermion spinor.

For massless spinors, there are only two non-zero invariant products which
are defined as follows,
\begin{equation}
s(p,q)= \ubar_+(p) u_-(q)=-s(q,p) \; ,
t(p,q)= \ubar_-(p) u_+(q)=\left [ s(q,p) \right ] ^* \; .
\end{equation}
In fact, it is enough to derive the expression of $s$ by using Eqs.(2.1)-(2.3).
We obtain
\begin{eqnarray}
s(p,q)&=&\ubar_-(k_0) \psla \qsla \ksla_1 \; u_-(k_0)/
\sqrt{4 (k_0 \cdot p)  (k_0 \cdot q)}\nonumber\\
  &=& Tr[\psla \qsla \ksla _1\ksla _0 \; \omega _+]/
\sqrt{4 (k_0 \cdot p) (k_0 \cdot q)} \;.
\end{eqnarray}
The expression for $t(p,q)$
can be obtained from the second equation in Eq.(2.4).
To calculation the invariant quantity $s(p,q)$, we can choose $k_0$ and
$k_1$ to be, for example,
\begin{equation}
k_0=(1,1,0,0) \; , \; k_1=(0,0,1,0) \; .
\end{equation}
With the help of Eqs.(2.5) and (2.6), $s(p,q)$ are given by
\begin{equation}
s(p,q)=(p^y + ip^z) \left [ {{q^0-q^x} \over {p^0-p^x}} \right ]^{1/2}
-(q^y + iq^z) \left [ {{p^0-p^x} \over {q^0-q^x}} \right ] ^{1/2}\;.
\end{equation}

Using Eqs.(2.1) and (2.4), we can derive the following useful formulae:
\begin{mathletters}
\begin{eqnarray}
\g^\m u_{\pm}(p) \ubar_{\pm}(q) \g_\m &=& -2 u_{\mp}(q) \ubar_{\mp}(p) \;,\\
\g^\m u_+(p) \ubar_-(q) \g_\m &=&2 \omega_- t(q,p)\;,\\
\g^\m u_-(p) \ubar_+(q) \g_\m &=&2 \omega_+ s(q,p)\;.
\end{eqnarray}
\end{mathletters}
Therefore, we can express any amplitude with external massless
fermions in terms of the invariant quantities $s$ and $t$.
For more general applications of the helicity amplitude technique involving
massive particles, the reader is recommended to read Ref.
\cite{5}.  For the purpose of this paper, however,
the above preliminary introduction is sufficient.

\subsection{The amplitudes of \eq}
In this section, we will compute the helicity amplitudes for the process
\eq\ .  The Feynman diagrams are shown in Fig. 1.
Using the Feynman Rules given in Ref \cite{1}, the corresponding amplitudes are
given by
\begin{mathletters}
\begin{eqnarray}
\FL
Amp(a)=\left ( {{g_{3l}} \over {\sqrt 2}} \right ) ^2 e^2 Q_q
     & &{{-1} \over {(p_2-p_4)^2}} {{-1} \over
{(p_5+p_6)^2-M_X^2+iM_X\Gamma_X}}
\nonumber\\
       & &\times {{1} \over {(p_3+p_5+p_6)^2}} M(a) \; ,
\end{eqnarray}
\begin{eqnarray}
Amp(b)=\left ( {{g_{3l}} \over {\sqrt 2}} \right ) ^2 e^2 Q_q
     & &{{-1} \over {(p_2-p_4)^2}} {{-1} \over
{(p_5+p_6)^2-M_X^2+iM_X\Gamma_X}}
\nonumber\\
       & &\times {{1} \over {(-p_1+p_5+p_6)^2}} M(b) \; ,
\end{eqnarray}
\begin{eqnarray}
Amp(c)=2\left ( {{g_{3l}} \over {\sqrt 2}} \right ) ^2 e^2 Q_q
    & &{{-1} \over {(p_2-p_4)^2}} {{-1} \over {(p_1-p_3)^2-M_X^2+iM_X\Gamma_X}}
\nonumber\\
       & &\times {{-1} \over {(p_5-p_6)^2-M_X^2+iM_X\Gamma_X}} M(c) \; ,
\end{eqnarray}
\begin{eqnarray}
Amp(d)=\left ( {{g_{3l}} \over {\sqrt 2}} \right ) ^2 e^2 Q_q
     & &{{-1} \over {(p_2-p_4)^2}} {{-1} \over
{(p_1-p_3)^2-M_X^2+iM_X\Gamma_X}}
\nonumber\\
     & &\times {{1} \over {(p_1-p_3-p_5)^2}} M(d) \; ,
\end{eqnarray}
\begin{eqnarray}
Amp(e)=\left ( {{g_{3l}} \over {\sqrt 2}} \right ) ^2 e^2 Q_q
    & & {{-1} \over {(p_2-p_4)^2}} {{-1} \over
{(p_1-p_3)^2-M_X^2+iM_X\Gamma_X}}
\nonumber\\
      & & \times{{1} \over {(p_1-p_3-p_6)^2}} M(e) \; ,
\end{eqnarray}
\end{mathletters}
where
\begin{mathletters}
\begin{equation}
M(a)=\ubar (p_4) \g _\a u(p_2) \ubar (p_6) \g_\m \g_5 C \ubar ^T(p_5)
     v^T(p_3) C \g^\m \g_5 (\psla _3 + \psla _5 + \psla _6) \g^a u(p_1)\; ,
\end{equation}
\begin{equation}
M(b)=\ubar (p_4) \g _\a u(p_2) \ubar (p_6) \g_\m \g_5 C \ubar ^T(p_5)\;
     v^T(p_3) C \g^\a (-\psla _1 + \psla _5 + \psla _6) \g^\m \g_5 u(p_1)\; ,
\end{equation}
\begin{eqnarray}
M(c)&&=\ubar (p_4) \g ^\a u(p_2) \ubar (p_6) \g^\m \g_5 C \ubar ^T(p_5)
        v^T(p_3) C \g^\b \g_5 u(p_1) \nonumber\\
&&\times \left [
	(p_2 - p_4 + p_5 + p_6)_\b {g_{\m\a}}
	+(-p_5 - p_6 - p_1 + p_3)_\a {g_{\m\b}}
        \right. \nonumber\\
&& \left.
	+(p_1 - p_3 - p_2 + p_4)_\m {g_{\a\b}}
   \right ] \; ,
\end{eqnarray}
\begin{equation}
M(d)=\ubar (p_4) \g_\a u(p_2) \ubar (p_6) \g^\a (\psla_1 - \psla_3 - \psla_5)
     \g^\m \g_5 C \ubar ^T (p_5) v^T(p_3) C \g_\m \g_5 u(p_1) \; ,
\end{equation}
\FL
\begin{equation}
M(e)=\ubar (p_4) \g_\a u(p_2) \ubar (p_6)\g^\m \g_5
     (\psla_1 - \psla_3 - \psla_6)
     \g^\a C \ubar ^T (p_5) v^T(p_3) C \g_\m \g_5 u(p_1) \; .
\end{equation}
\end{mathletters}
$\Gamma_X$ is the total width of $X^{--}$ which decays into $e^-e^-,\mu^-\mu^-$
and $\tau^-\tau^-$ democratically.
After some Dirac matrix manipulation, Eq.(2.10c) can be rewritten as
\begin{equation}
M(c)=-M(a)-M(c) \;.
\end{equation}

For massless spinors, we can replace $v(p)$ by $u(p)$.
Therefore, we can decompose $M(a)-(e)$ into various helicities as follows:
\begin{mathletters}
\begin{eqnarray}
M_{\pm\pm\pm}(a)=& &
\left [
\begin{array}{c}
\ubar_+ (p_4) \g_\a u_+(p_2)\\ \ubar_- (p_4) \g_\a u_-(p_2)
\end{array}
\right ]
\left [
\begin{array}{c}
\ubar_+ (p_6) \g_\m u_+(p_5)\\ -\ubar_- (p_6) \g_\a u_-(p_5)
\end{array}
\right ]\nonumber\\
& &\times
\left [
\begin{array}{c}
\ubar_+ (p_3) \g^\m (\psla_3 + \psla_5 + \psla_6) \g^\a u_+(p_1)\\
-\ubar_- (p_3) \g^\m (\psla_3 + \psla_5 + \psla_6) \g^\a u_-(p_1)
\end{array}
\right ]\; ,
\end{eqnarray}
\begin{eqnarray}
M_{\pm\pm\pm}(b)=& &
\left [
\begin{array}{c}
\ubar_+ (p_4) \g_\a u_+(p_2)\\ \ubar_- (p_4) \g_\a u_-(p_2)
\end{array}
\right ]
\left [
\begin{array}{c}
\ubar_+ (p_6) \g_\m u_+(p_5)\\ -\ubar_- (p_6) \g_\m u_-(p_5)
\end{array}
\right ]\nonumber\\
& &\times
\left [
\begin{array}{c}
\ubar_+ (p_3) \g^\a (-\psla_1 + \psla_5 + \psla_6) \g^\m u_+(p_1)\\
-\ubar_- (p_3) \g^\a (-\psla_1 + \psla_5 + \psla_6) \g^\m u_-(p_1)
\end{array}
\right ]\; ,
\end{eqnarray}
\begin{equation}
M_{\pm\pm\pm}(c)=-M_{\pm\pm\pm}(a)-M_{\pm\pm\pm}(b)\; ,
\end{equation}
\begin{eqnarray}
M_{\pm\pm\pm}(d)=& &
\left [
\begin{array}{c}
\ubar_+ (p_4) \g_\a u_+(p_2)\\ \ubar_- (p_4) \g_\a u_-(p_2)
\end{array}
\right ]
\left [
\begin{array}{c}
\ubar_+ (p_6) \g^\a (\psla_1 - \psla_3 - \psla_5) \g^\m u_+(p_5)\\
-\ubar_- (p_6) \g^\a (\psla_1 - \psla_3 - \psla_5) \g^\m u_-(p_5)\\
\end{array}
\right ]\nonumber\\
& &\times
\left [
\begin{array}{c}
\ubar_+ (p_3) \g_\m u_+(p_1)\\ -\ubar_- (p_3) \g_\m u_-(p_1)
\end{array}
\right ]\; ,
\end{eqnarray}
\begin{eqnarray}
M_{\pm\pm\pm}(e)=& &
\left [
\begin{array}{c}
\ubar_+ (p_4) \g_\a u_+(p_2)\\ \ubar_- (p_4) \g_\a u_-(p_2)
\end{array}
\right ]
\left [
\begin{array}{c}
\ubar_+ (p_6) \g^\a (\psla_1 - \psla_3 - \psla_6) \g^\m u_+(p_5)\\
-\ubar_- (p_6) \g^\a (\psla_1 - \psla_3 - \psla_6) \g^\m u_-(p_5)\\
\end{array}
\right ]\nonumber\\
& &\times
\left [
\begin{array}{c}
\ubar_+ (p_3) \g_\m u_+(p_1)\\ -\ubar_- (p_3) \g_\m u_-(p_1)
\end{array}
\right ]\;.
\end{eqnarray}
\end{mathletters}
Since $M_{\lambda_1 \lambda_2 \lambda_3} =
M^*_{-\lambda_1 -\lambda_2 -\lambda_3}$,
we need calculate only the helicity amplitudes,$M_{+\pm\pm}(a)-(e)$,
explicitly in terms of $s$ and $t$.  The results are given in Table 1.

\section{cross-sections}

Since the violation of $L_e$ and $L_\mu$ conservation is clearly evidenced by
the processes \eq \ and \ee \ , it is totally free of the minimal standard
model background.  Before we proceed, let us justify neglecting the Feynman
diagrams in which the photon of Figs. 1(a)-(e) is replaced by a $Z$-boson.
Aside from the suppression due to the mass in the $Z$ propagator, the axial
vector couplings of electron and Z-boson do not contribute in this process
because of the Fermi statistics, see Ref. \cite{1}.  Only the vector coupling
of $Z$ contributes, but it is proportional to $g_v = ({1\over4} -
sin^2\theta_W)
\simeq 0.02$. The three boson coupling of $X^{--} - X^{++} - Z$ is also
proportional to $g_v$ from the group theory.
Therefore, the $Z$-boson contributes at most $0.5 \%$  to the processes and
it can be safely neglected.

\subsection{$e^-p$ colliders}

To evaluate the production cross-section for the process \ep \  in the
electron-proton colliders, we use EHLQ \cite{6} parton structure functions
(set 1), $F_q(x)$ for quark q.
Hence the production cross-section for the process is given by
\begin{equation}
\sigma(M_X) = \int_0^1 dx \sum_q F_q(x,Q^2)
\hat{\sigma} (\sqrt {\hat{s}} = x s, M_X)\;,
\end{equation}
where $\hat{\sigma}$ is the elementary cross-section of the process \eq \ ;
$x$ is the fractional momentum of the proton carried by the quark $q$, hence
$\sqrt {\hat{s}}$ is the center of mass energy available for \eq \ .
$Q^2$, defined to be $-(p_2-p_4)^2$, is the scale for the structure functions
for quarks.  The result for $\sigma (M_X)$ are shown in Fig. 2 for the cases
$\sqrt s = 314 GeV$ (HERA) and $1790GeV$ (LEPII-LHC).

For HERA, the planned luminosity is $1.6 \times 10^{31} cm^{-2}s^{-1}$
giving an annual integrated luminosity of $0.5 fb^{-1}yr^{-1}$. From Fig. 2,
we see that there will be less than one event per year if the mass of the
dilepton is heavier than $120 GeV$.  The situation become hopeless for $M_X
> 150 GeV$ without an upgrade in energy and/or luminosity.  For example,
an up-grade in center of mass energy up to $400 GeV$ will allow, for the same
luminoisity, discovery of dileptons up to about $200 GeV$.  We thus conclude,
given the mass bounds mentioned in the introduction, that the chance of HERA
discovering such a dilepton state is very marginal.

At LEPII-LHC with $\sqrt s = 1790 GeV$ the prospects for dilepton discovery are
far better.  The expected luminosity
is about $2 \times 10^{32} cm^{-2} s^{-1}$ and hence annual integrated
luminosity $6 fb^{-1}yr^{-1}$.  Requiring at least 2 events per year for
\ep \ , we can detect $M_X$ up to $650 GeV$.

\subsection{$e^+e^-$ colliders}

At an $e^+ e^-$ collider, dilepton signatures include \ee \ .
This calculation is quite similar to \eq \ described above and we include also
the charge-conjugation of the corresponding Feynman diagrams.
We have computed the result for the center-of-mass
energies $\sqrt s = 200 GeV$(LEPII), $500 GeV$ and $1000 GeV$ (possible NLC
energies).  The results are displayed in Fig. 3.
Requiring at least 2 events per year, we can
detect $M_X$ up to 180,450 and 950 GeV in $e^+e^-$ colliders with energies
$\sqrt{s} = 200, 500$ and $1000 GeV$ assuming the integrated luminosities
\cite{4} to be $0.5,3$ and $30$ $fb^{-1} yr^{-1}$ respectively.

The amplitude-squared for the real production of the dilepton is given in
the Appendix A.  The production cross-sections are also calculated and compared
with the curves in Figs. (2) and (3).  We find that the contribution from the
Feynman diagrams Figs. 1 (d) and (e) are at most $10\%$ relative to that of
other diagrams in a wide range of dilepton mass $M_X$ except at the high
values of $M_X$ in which the curves have longer tails.
Therefore, it is important
to include Figs. 1 (d) and (e) in our calculation in order to provide a better
estimation for the maximum $M_X$ being probed in high energy colliders.

\section{conclusion}

We have considered a direct search for doubly-charged dilepton
$X^{--}(X^{++})$ by lepton-number violating processes in $e^- p$ and $e^+ e^-$
colliders.  The mass of $X^{--}$ ranging from $100$ to $1000 GeV$ is expected
from the theory of $SU(15)$.  The striking signature for a dilepton gauge boson
is \ep \ in an $e^- p$ collider and \ee \ in an $e^+ e^-$ collider.  The chance
of discovering a dilepton at HERA is very marginal unless $M_X$ is less than
$150 GeV$.  The direct discovery of such a dilepton state depends on future
colliders such as LEPII-LHC and NLC at which interesting mass ranges will be
explored.

\acknowledgments

We thank S. L. Glashow for a useful suggestion.  This work was supported in
part by the U. S. Department of Energy under Grant No. DE-FG05-85ER-40219.

\figure{Feynman diagrams for $e^-q\rightarrow e^+\mu ^- \mu^-q$}
\figure{Cross-sections for the process \ep \  with $Q^2 > 25 GeV^2$
at the center-of-mass energies
$ \sqrt{s}=314 GeV$ (solid line) and $ \sqrt{s}=1790 GeV$ (dashed line)}
\figure{Cross-sections for the process $e^+e^-\rightarrow 2e^- 2\mu^+$
or $2e^+2\mu^-$ with $Q^2 > 25 GeV^2$
at the center-of-mass energies
$ \sqrt{s}=200 GeV$ (solid line), $ \sqrt{s}=500 GeV$ (dashed line) and
$\sqrt{s}=1000 GeV$ (dotted line)}

\mediumtext
\begin{table}
\caption{Helicity amplitudes for the Feynman diagrams shown in Fig. 1}
\begin{tabular}{ccccc}
$M$ & $+++$ & $++-$ &$+-+$ &$+--$ \\
 \hline
 $M(a)$
 &$4t(p_2,p_1)s(p_6,p_3)$
 &$-4t(p_5,p_3)s(p_4,p_1)$
 &$-4t(p_2,p_1)s(p_5,p_3)$
 &$4t(p_6,p_3)s(p_4,p_1)$ \hfill \\
  &$\times[t(p_5,p_3)s(p_4,p_3)$
  &$\times[t(p_2,p_3)s(p_6,p_3)$
  &$\times[t(p_6,p_3)s(p_4,p_3)$
  &$\times[t(p_2,p_3)s(p_5,p_3)$ \hfill \\
    &$+t(p_5,p_6)s(p_4,p_6)]$
    &$+t(p_2,p_5)s(p_6,p_5)]$
    &$+t(p_6,p_5)s(p_4,p_5)]$
    &$+t(p_2,p_6)s(p_5,p_6)]$ \hfil\\
\hline
 $M(b)$
 &$4t(p_5,p_1)s(p_4,p_3)$
 &$-4t(p_2,p_3)s(p_6,p_1)$
 &$-4t(p_6,p_1)s(p_4,p_3)$
 &$4t(p_2,p_3)s(p_5,p_1)$ \hfill \\
  &$\times[t(p_2,p_5)s(p_6,p_5)$
  &$\times[t(p_5,p_6)s(p_4,p_6)$
  &$\times[t(p_2,p_6)s(p_5,p_6)$
  &$\times[t(p_6,p_5)s(p_4,p_5)$ \hfill \\
    &$-t(p_2,p_1)s(p_6,p_1)]$
    &$-t(p_5,p_1)s(p_4,p_1)]$
    &$-t(p_2,p_1)s(p_5,p_1)]$
    &$-t(p_6,p_1)s(p_4,p_1)]$ \hfil\\
\hline
\multicolumn{1}{c}{$M(c)$} &\multicolumn{4}{c}{$-M(a)-M(b)$}\\
\hline
 $M(d)$
 &$4t(p_1,p_5)s(p_4,p_6)$
 &$-4t(p_3,p_5)s(p_4,p_6)$
 &$-4t(p_2,p_6)s(p_3,p_5)$
 &$4t(p_2,p_6)s(p_1,p_5)$ \hfill \\
  &$\times[t(p_2,p_1)s(p_3,p_1)$
  &$\times[-t(p_2,p_3)s(p_1,p_3)$
  &$\times[-t(p_1,p_3)s(p_4,p_3)$
  &$\times[t(p_3,p_1)s(p_4,p_1)$ \hfill \\
    &$-t(p_2,p_5)s(p_3,p_5)]$
    &$-t(p_2,p_5)s(p_1,p_5)]$
    &$-t(p_1,p_5)s(p_4,p_5)]$
    &$-t(p_3,p_5)s(p_4,p_5)]$ \hfil\\
\hline
 $M(e)$
 &$4t(p_2,p_5)s(p_3,p_6)$
 &$-4t(p_2,p_5)s(p_1,p_6)$
 &$-4t(p_1,p_6)s(p_4,p_5)$
 &$4t(p_3,p_6)s(p_4,p_5)$ \hfill \\
  &$\times[-t(p_1,p_3)s(p_4,p_3)$
  &$\times[t(p_3,p_1)s(p_4,p_1)$
  &$\times[t(p_2,p_1)s(p_3,p_1)$
  &$\times[-t(p_2,p_3)s(p_1,p_3)$ \hfill \\
    &$-t(p_1,p_6)s(p_4,p_6)]$
    &$-t(p_3,p_6)s(p_4,p_6)]$
    &$-t(p_2,p_6)s(p_3,p_6)]$
    &$-t(p_2,p_6)s(p_1,p_6)]$ \hfil\\
\end{tabular}
\end{table}
\begin{appendix}\\
In this appendix, we will present the calculation of the real production
of dilepton $X^{--}$.
If there is sufficient center-of-mass energy and the dilepton is light enough,
it will be possible to produce a real dilepton in the final state.  This limit
of light dilepton mass provides, in any case, a useful check on all the
calculations given in the main text.  Clearly, for a light dilepton, the
calculation based on the Feynman diagrams given in Figure 1 with a Breit-
Wigner form of the dilepton propagator should agree with a real dilepton
calculation using three-body (rather than four-body) phase space.
It is because of the fact that Figs. 1(a)-(c) are dominant over
Figs. 1 (d) and (e).  The success
of this comparison gives us confidence that the four-body phase space
calculation in the main text is reliable.
Only the first three diagrams in Fig. 1 are relevant.
The amplitude is
\begin{equation}
Amp=Q_qe^2{g_{3l}\over{\sqrt 2}}\epsilon^{\mu}(p)a_{\mu \alpha}b^\alpha
{1 \over{(p_2-p_4)^2}}\; ,
\end{equation}
where
\begin{eqnarray}
a_{\mu \alpha} =&& v^T(p_3)C\gamma_\mu \gamma_5
    {{\psla_3+\psla}\over{(p_3+p)^2}} \gamma_a u(p_1)
    + v^T(p_3)C\gamma_\alpha
     {{-\psla_1+\psla}\over{(-p_1+p)^2}} \gamma_\mu \gamma_5 u(p_1)\nonumber\\
    && + v^T(p_3)C\gamma_\beta \gamma_5 u(p_1) {{-2}\over{(p_1-p_3)^2-M_X^2}}
    \left [ (p_2-p_4+p)_\beta g_{\mu\alpha}
    \right.\nonumber\\
    &&
    \left.  + (-p-p_1+p_3)_\alpha g_{\mu\beta}
        +(p_1-p_3-p_2+p_4)_\mu g_{\alpha\beta}\right ] ,
\end{eqnarray}
and
\begin{equation}
b^\alpha=\ubar(p_4)\gamma^\alpha u(p_2)\;,
\end{equation}
where $p_1$ and $p_3$ are the momenta for the electron and positron;
$p_2$ and $p_4$ are the momenta for the initial and final quarks; $p$
and $\epsilon^\mu(p)$ are the momenta and polarization vector for the
dilepton $X^{--}$.  Here $eQ_q$ is the quark electric charge and
$g_{3l}/\sqrt{2}$ is the coupling constant for the $X^{++}-e-e$ interaction.
We have neglected the unimportant width of $X^{--}$ in the propagator.
Notice that $(p_2-p_4)^\mu a_{\mu\alpha}=0$ because of electromagnetic
gauge invariance. Using momentum conservation and Dirac algebra (see Eq.
(2.11) in the text), $a_{\mu \alpha}$ in the Eq. (2)
can be rewritten as
\begin{eqnarray}
a_{\mu\alpha}\;&=& \left [ {1\over{(p_3+p)^2}}
                  +{2\over{(p_1-p_3)^2-M_X^2}} \right ]
          v^T(p_3)C\gamma_\mu(\psla_3+\psla)\gamma_\alpha\gamma_5u(p_1)
\nonumber\\
   &+&\left [ {1\over{(-p_1+p)^2}}+{2\over{(p_1-p_3)^2-M_X^2}}\right ]
         v^T(p_3)C\gamma_\alpha(-\psla_1+\psla)\gamma_\mu\gamma_5u(p_1)\;.
\end{eqnarray}
$p^\mu a_{\mu\alpha}$ is not zero because the dilepton is not coupled to
a conserved current; in fact it is given explicitly by
\begin{eqnarray}
p^\mu a_{\mu\alpha}=2{{(p_2-p_4)^2}\over{(p_1-p_3)^2-M_X^2}}
       v^T(p_3)C \gamma_\alpha\gamma_5u(p_1)\;.
\label{eq:2}
\end{eqnarray}
Using the polarization sum $\sum \epsilon^\mu(p)\epsilon^\nu(p)=
-g^{\mu\nu}+p^\mu p^\nu/M_X^2$, the amplitude-squared is given by
\begin{equation}
|Amp|^2=Q_qe^4\left ( {{g_{3l}}\over\sqrt 2}\right ) ^2
{1\over{(p_2-p_4)^4}}
\left ( -g_{\mu\nu}+{{p_\mu p_\nu}\over{M_X^2}}\right ) a_{\mu\alpha}
a^\ast_{\nu\beta}b^\alpha b^{\beta\ast} \;.
\end{equation}
Therefore $|Amp|^2$, with the help of Eq.(5), can be calculated to be
\begin{eqnarray}
|Amp|^2\;&&= Q_q^2e^4\left ( {g_{3l}\over{\sqrt 2}}\right ) ^2
  {{64}\over{(p_2-p_4)^4}}
\left [
\left({1\over(p_3+p)^2}+{2\over{(p_1-p_3)^2-M_X^2}}\right) ^2
\right.
\nonumber\\
    &&\times \left [
      2 \; p_3.p \;(p_1.p_2 \; p.p_4+p_1.p_4 \; p.p_2)
     -M_X^2 \; (p_1.p_2 \; p_3.p_4+p_1.p_4 \; p_3.p_2)
       \right ] \nonumber\\
    && +\left ( {1\over{(-p_1+p)^2}}+ {2\over{(p_1-p_3)^2-M_X^2}}\right )^2
\nonumber\\
    && \times
\left [2 \; p_1.p \;(p_3.p_2 \; p.p_4+p_3.p_4 \; p.p_2)
     -M_X^2 \; (p_1.p_2 \; p_3.p_4+p_1.p_4 \; p_3.p_2)
\right ]
\nonumber\\
    && + 2
\left ( {1\over{(p_3+p)^2}}+{2\over{(p_1-p_3)^2-M_X^2}} \right )
\left ( {1\over{(-p_1+p)^2}}+{2\over{(p_1-p_3)^2-M_X^2}}\right )
\nonumber\\
    && \times
\bigg [
 -M_X^2 \; p_1.p_3 \; p_2.p_4
       + 2\; p_1.p \; p_3.p_2 \; p_3.p_4-2 \; p_3.p \; p_1.p_2 \; p_1.p_4
\nonumber\\
    &&    +(2 \; p_1.p_3+p_1.p-p_3.p)(p_1.p_2 \; p_3.p_4+p_1.p_4 \; p_3.p_2)
\nonumber\\
    &&
      +p_1.p_3 \;(p_1.p_2 \; p.p_4+p_1.p_4 \; p.p_2
         -p.p_2 \; p_3.p_4-p.p_4 \; p_3.p_2)
\bigg ]
\nonumber\\
    &&
\left.
+2\left ( {1\over{(p_1-p_3)^2-M_X^2}}\right ) ^2{{(p_2-p_4)^4} \over{M_X^2}}
      \; (p_1.p_2\;p_3.p_4+p_1.p_4\;p_3.p_2)
\right ] \; .
\end{eqnarray}
The above equation is used to calculate the production of a real dilepton in
the $e^-p$ and $e^+e^-$ collders.  We then compared this result with Figs. (2)
and (3).  We find agreement for light dilepton mass with the curves in
Figs. (2) and (3) and that, as expected, the full calculation allowing a
virtual
dilepton gives an extra contribution in the tail of high $M_X$ values.
\end{appendix}
%\newpage
%\next page
%\newpage
%\vfill\eject
%\input epsf
%\centerline{\epsffile{ep.ps}}
%\centerline{Figure 2}
%\vfill\eject
%\input epsf
%\centerline{\epsffile{nlc.ps}}
%\centerline{Figure 3}
\end{document}